# Statistics of optical intraday variability in a complete sample of radio-selected BL Lac objects[*, **]

J. Heidt[1] and S. J. Wagner[1]

Landessternwarte Heidelberg, Königstuhl,
69117 Heidelberg, Germany

**Abstract.** We present a study of the intraday variability behaviour of a complete sample of radio-selected BL Lac objects taken from the 1 Jy catalogue. In 28 out of 34 BL Lac objects (82%) we detected intraday variability. 7 objects were observed during several campaigns. None of them changed its variability behaviour. The duty cycle in radio-loud BL Lac objects is very high, i.e. at least 0.8.

The typical peak-to-peak amplitudes of the variability were 30%. By means of structure function and autocorrelation analyses we investigated the typical time-scales of the variability and determined the activity of the BL Lac objects. In 21 of the variable BL Lac objects we were able to measure a typical time-scale, which lies in the range between 0.5 and 3 days.

In general, the results can be explained by the standard model, where shocks are propagating down a relativistic jet. However, in the two most densely sampled observations we found variability characteristics, which cannot easily be explained by the standard model alone, demonstrating that at least in these objects alternative models should be taken into account. Nevertheless, all models, which are able to explain variability on time-scales of days must be able to consider the high duty cycle of these BL Lac objects.

**Key words:** Galaxies: active – BL Lacertae objects: general – intraday variability – Methods: statistical

## 1. Introduction

BL Lac objects are one of the most puzzling classes of the active galactic nuclei (AGN). They are characterized by their strong and rapid variability of the total and polarized flux and absent or faint lines in their spectra. The AGN catalogue of Véron-Cetty and Véron (1991) contains 162 BL Lac objects, the revised and updated catalog of quasi-stellar objects by Hewitt & Burbidge (1993) contains about 90 BL Lac objects. It is expected that their number will grow to more than 1000 after the identification of the x-ray sources found in the ROSAT All Sky Survey (Voges 1992).

Optical variability measurements are one tool to explore the physics of these objects. Nevertheless, such variability studies were carried out rather inhomogeneous in the past. Some are included in long-term monitoring programs of AGN (e.g. Pica et al. 1988; Webb et al. 1988; Borgeest et al. 1991; Jannuzi et al. 1993). A few are studied extensively over a wide range of time-scales (e.g. BL Lac (Bertaud et al. 1969; Racine 1970; Epstein et al. 1972; Moore et al. 1982; Brown et al. 1989) or OJ 287 (Takalo, 1994 and references therein)).

In 1989 a simultaneous multi-frequency campaign (radio-optical) was carried out in order to search for correlated variability in BL Lac objects and quasars from the S5 survey (Kühr et al. 1981a). In 3 of 4 BL Lac objects intraday variability was found both in the optical and the radio domain (Wagner et al. 1990). More densely sampled follow-up studies of these objects confirmed this variability behaviour (Quirrenbach et al. 1991; Wagner 1991; Wagner et al. 1993; Wagner et al. submitted; Bock 1994). Whereas the intraday variability in the radio domain can (partly) be attributed to refractive interstellar scattering (Fiedler et al. 1987)) short-term variability in the optical domain is most likely intrinsic in nature (Wagner 1992). Recent investigations have shown that intraday variability in the radio domain is a characteristic property of flat-spectrum radio sources (which form the parent group of the classical radio-selected BL Lac objects) (e.g. Witzel et al. 1986;

et al. 1992). However, it is not clear, whether variability on time-scales of days in the optical domain is typical for the BL Lac class in general or a characteristic feature of a few specific BL Lac objects. To investigate this issue statistical investigations of the variability behaviour on time-scales of days of a complete sample of BL Lac objects are required.

Due to rather inhomogeneous identification programs and an ongoing debate in the literature about the definition of a BL Lac and its distinguishing criteria from other AGN and the fact that only a small number of them is known it is not easy to create a complete sample of BL Lac objects. Burbidge & Hewitt (1991) proposed to confine the BL Lac term only to those objects, whose spectra are typical for non-thermal sources lying in bright elliptical galaxies. Valtaoja et al. (1991) suggested to distinguish between low redshift ($z < 0.3$) and high redshift ($z > 0.3$) BL Lac objects with different parent populations, respectively. Additionally, it has been proposed to distinguish between radio-selected (RBL) and x-ray selected (XBL) BL Lac objects (Garilli et al. 1990). They differ in their overall electromagnetic spectra and it is often assumed that the XBL are not as strongly variable and not as highly polarized as the RBL (Stocke et al. 1985).

To our knowledge the only complete and well defined sample of BL Lac objects is the 1 Jy sample of radio-selected BL Lac objects (Stickel et al. 1991). The intraday variability behaviour of the BL Lac objects from this sample is the topic of the present paper. We determine the properties of variability on time-scales from a few hours to about one week. The variability behaviour of XBLs and the relationship between the two classes of BL Lac objects within the Unified Scheme will be the subject of further papers (Heidt et al. in preparation). The variability behaviour on even shorter time-scales (minutes to hours) were studied in a few selected objects (previously known to be very active) by Miller et al. (1989), Carini (1990), Carini et al. (1990, 1992) and Carini & Miller (1992).

This paper is organized as follows: In chapter 2 the 1 Jy sample is briefly described, in chapter 3 the observations and the data reduction are summarized. Chapter 4 contains the basic results, the statistical analysis is presented in chapter 5. In chapter 6 we discuss the results of the statistical analysis and draw conclusions in chapter 7. Throughout the paper $H_0 = 50\ km\ s^{-1}\ Mpc^{-1}$ and $q_0 = 0$ is assumed.

## 2. The 1 Jy sample

The 1 Jy sample of BL Lac objects was extracted by Stickel et al. (1991) from the 1 Jy catalogue (Kühr et al. 1981b). The revised compilation (Stickel et al. 1994) contains 527 sources with a flux-density $S_{5GHz} \geq 1Jy$ and covers 9.811sr of the sky avoiding the galactic plane ($|b| \leq 10^\circ$). The 1 Jy sample of BL Lac objects contains ing criteria: a) flat or inverted radio spectra ($\alpha_{11-6cm} \geq -0.5$ ($S_\nu \sim \nu^\alpha$)), b) equivalent widths $\leq 5$Å (in rest frame) of the emission lines and c) $m_V \leq 20$ on sky plates. In 26 of the 34 objects redshifts could be measured from emission lines or absorption lines from the host galaxy. Of the remaining 8 BL Lac objects 4 displayed intergalactic MgII absorption line systems in their spectra. The redshift of those are given as lower limits in Stickel et al. (1991). Of the remaining 4 BL Lac objects neither emission nor absorption lines could be measured so far. Since they appear stellar on deep images (Stickel et al. 1993), we assume a lower limit to the redshift of these objects of $z = 0.3$ throughout the paper. Thus the redshift range, which is covered by the 1 Jy sample of BL Lac objects lies between z = 0.033 and 1.048.

## 3. Observations and data reduction

The observations were carried out during 10 observing runs between June 1990 and September 1993 with telescopes ranging in size from 0.7m up to 3.5m at the observatory in Heidelberg, at the Calar Alto observatory in Spain and at the ESO in Chile. The telescopes were equipped with a CCD and an R filter in order to perform relative photometry. An overview about the observing campaigns is given in table 1.

**Table 1.** Observing journal

| Observatory | Period |
| --- | --- |
| ESO, Danish 1.5m | June 1990 |
| Calar Alto, 3.5m | August 1990 |
| Heidelberg, 0.7m | August 1990 - March 1991 |
| Calar Alto, 2.2m | January 1991 |
| Heidelberg, 0.7m | October 1991 |
| Calar Alto, 1.2m | January 1992 |
| ESO, Danish 1.5m | September 1992 |
| Calar Alto, 2.2m | February 1993 |
| Calar Alto, 2.2m | May 1993 |
| ESO, Danish 1.5m | September 1993 |

In order to create a homogeneous dataset we aimed in observing each BL Lac object during seven consecutive nights with an average sampling rate of 2 hours. In practice, we observed a BL Lac object typically 21 times during 7 nights. However, due to various circumstances (e.g. weather, technical problems, average brightness of the BL Lac object, etc.) the dispersion in the sampling rates and number of observations is rather broad. 4 BL Lac objects (PKS 0735+178, OJ 287, S4 1749+701 and 3C 371) were

0716+714 and S4 0954+658) were observed more than 80 times during 8 nights (c.f. table 2 and figure 4).

The integration times varied between 2 and 30 minutes depending on the brightness of the BL Lac object, the observing conditions and the size of the telescope. They were chosen such that the signal to noise in the central pixel of the BL Lac object was as high as possible but the count rate approximately 30% below the saturation limit or the non-linearity limit and the count rates of the comparison stars varied no more than a factor of two within each campaign.

The CCD frames were corrected for bias by taking several bias frames each night and averaging them over the entire campaign. As most CCDs had a variable bias, we scaled the bias during the observations via the overscan. This procedure was possible for all CCDs except the CCD, which we used for the observations in Heidelberg. In that case we scaled the bias according to the average bias for each night. Each CCD was checked for dark current and the CCD frames were corrected for if necessary. The pixel-to-pixel variations at the CCD were removed by taking twilight flatfields and averaging them. In a few cases two or more flatfields for removing the pixel-to-pixel variations were necessary e.g. when there were non-stationary features at the CCD frames (dust particles at the filter or the entrance window of the CCD). During the observations in January 1992 and from August 1990 - March 1991, which were partly carried out during bright time, scattered light from the moon added extra light on the CCD. This resulted in a gradient up to several hundred ADU across the CCD. These feature were removed by fitting and subtracting a second-order polynomial.

In order to carry out relative photometry the count rates of the BL Lac object and 5 to 10 comparison stars were measured on each frame by simulated aperture photometry. We computed the normalized ratio for each pair of objects over the entire campaign to construct lightcurves. By inspecting the lightcurve of each pair, variable stars were found and rejected from further analysis. For the final analysis the lightcurves were used which included the BL Lac object and the brightest, non-saturated comparison star. The errors were estimated from the standard deviation (1 $\sigma$) of the lightcurve of two comparison stars as bright as or fainter than the BL Lac object. They are typically in the order of 0.8-2.5%, in worst cases 5% (corresponding to 0.06mag for a 19th magnitude object).

## 4. Variability statistics and amplitudes

In order to check how many BL Lac objects displayed variability during the observations we applied a $\chi^2$-test (Penston & Cannon 1970) to each lightcurve. We choose a confidence level of 99.5% as cutoff. The $\chi^2$-test was carried out twice. First we applied this test to all lightcurves. This gives us the fraction of the variable and the non-from several hours up to one week. Hence we subtracted trends on even longer time-scales by fitting a linear slope to each lightcurve of the variable objects. After subtraction of these slopes, the $\chi^2$-test was applied to the residuals.

28 out of the 34 objects (82%) showed variability on time-scales of days or less. 2 objects displayed strong long-term variability ( S5 0454+844 and OT 081), but in both objects we found superimposed short-term variability. In table 2 we list the 1 Jy sample, the average magnitude in the R band during the observations, the redshift, the number of observations, the maximum time separation between two observations and in column 6 the result of the $\chi^2$-test after subtracting the linear slope. A + sign denotes the variable and a - sign the non-variable BL Lac objects. In column 7 we list the variability amplitudes as defined below and the measurement error. Column 8 contains information about the observing period and telescope used. Some BL Lac objects differed in brightness by more than two magnitudes from the values given in Stickel et al. (1991) (e.g. B2 1308+328 or PKS 0537-441). We will use our observed average magnitudes for the further analysis. In figure 1 we show a typical lightcurve of a variable and a non-variable BL Lac object.

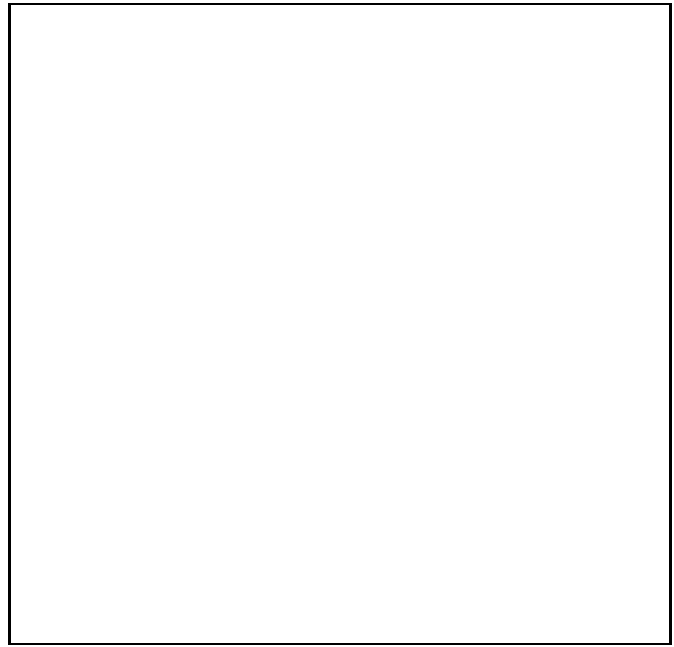

**Fig. 1.** Lightcurve of a variable (PKS 0537-441) and a non-variable BL Lac object (PKS 0118-272).

We checked, whether a lower confidence level, e.g. 90% as used by Penston & Cannon (1970) or 95% as used by Pica et al. (1988) would alter the distribution. We run through the procedure for both cutoff limits as outlined

**Table 2.** The 1 Jansky sample

| Object | $m_R$ | z | n | dt [days] | $\chi^2$ | Amplitude [%] | Date/Telescope |
|---|---|---|---|---|---|---|---|
| PKS 0048-097 | 16.5 |       | 45    | 6.38         | +     | 14.1(1.0)              | 09/92 (ESO,1.5) |
| PKS 0118-272 | 17.0 | 0.557 | 43    | 6.43         | -     | 5.3(1.5)               | 09/92 (ESO,1.5) |
| PKS 0139-097 | 17.0 | 0.501 | 12;10 | 4.03; 2.33   | - ; - | 11.7(3.2); 1.8(1.2)    | 09/93 (ESO,1.5); 01/92 (CA,1.2) |
| AO 0235+164  | 16.5 | 0.940 | 18    | 13.05        | +     | 41.8(1.5)              | 10/91 (Hd,0.7) |
| PKS 0426-380 | 18.0 | 1.030 | 27    | 6.26         | +     | 18.8(1.0)              | 09/92 (ESO,1.5) |
| S5 0454+844  | 19.0 | 0.112 | 13;13 | 10.01;9.77   | + ; + | 21.0(1.2); 16.0(3.2)   | 02/93 (CA,2.2); 01/92 (CA,1.2) |
| PKS 0537-441 | 13.5 | 0.896 | 20    | 5.23         | +     | 22.2(1.0)              | 09/92 (ESO,1.5) |
| S5 0716+714  | 13.0 |       | 249   | 11.37        | +     | 69.4(1.2)              | 01/91 (Hd,0.7 + CA,2.2) |
| PKS 0735+178 | 14.5 | 0.424 | 21    | 140.80       | +     | 65.7(2.2)              | 02/91 (Hd,0.7) |
| OJ 425       | 20.0 | 0.258 | 7     | 6.02         | +     | 8.8(1.5)               | 02/92 (CA,3.5) |
| PKS 0820+225 | 20.0 | 0.951 | 8     | 7.05         | +     | 8.5(1.4)               | 02/92 (CA,3.5) |
| PKS 0823+033 | 17.0 | 0.506 | 13    | 7.17         | -     | 4.0(1.3)               | 01/92 (CA,1.2) |
| OJ 448       | 19.0 | 0.548 | 14    | 9.13         | -     | 14.1(5.6)              | 01/92 (CA,1.2) |
| OJ 287       | 15.0 | 0.306 | 41    | 41.93        | +     | 48.5(1.2)              | 01/92 (Hd,0.7) |
| S4 0954+658  | 15.5 | 0.367 | 86    | 8.46         | +     | 76.3(1.7)              | 01/91 (Hd,0.7 + CA,2.2) |
| PKS 1144-379 | 16.5 | 1.048 | 19    | 4.07         | +     | 38.4(0.9)              | 06/90 (ESO,1.5) |
| B2 1147+245  | 16.5 |       | 10    | 6.97         | +     | 20.8(1.3)              | 02/93 (CA,2.2) |
| B2 1308+326  | 19.0 | 0.997 | 20    | 9.00         | +     | 12.6(2.3)              | 01/92 (CA,1.2) |
| OQ 530       | 14.5 | 0.152 | 21    | 5.18         | +     | 29.6(1.1)              | 05/93 (CA,2.2) |
| AP Lib       | 15.5 | 0.049 | 28    | 4.00         | +     | 5.7(0.5)               | 06/90 (ESO,1.5) |
| PKS 1519-273 | 18.5 |       | 5     | 4.97         | -     | 1.3(0.8)               | 05/93 (CA,2.2) |
| 4C 14.6      | 18.0 | 0.605 | 24    | 6.16         | +     | 15.9(1.1)              | 05/93 (CA,2.2) |
| MRK 501      | 14.0 | 0.033 | 8     | 13.88        | +     | 31.9(3.8)              | 08/90 (Hd,0.7) |
| OT 081       | 15.5 | 0.320 | 20    | 3.18         | +     | 51.6(0.8)              | 06/90 (ESO,1.5) |
| S4 1749+701  | 16.5 | 0.770 | 39;33 | 4.18; 76.81  | + ; + | 39.6(0.8); 68.0(1.1)   | 09/90 (Hd,0.7); 08/90 (CA,3.5) |
| S5 1803+784  | 16.5 | 0.684 | 37;23 | 4.12; 13.01  | + ; + | 26.7(0.8); 22.5(2.3)   | 10/91 (Hd,0.7); 08/90 (CA,3.5) |
| 3C 371       | 14.0 | 0.051 | 34;41 | 4.06; 75.88  | + ; + | 19.2(0.8); 18.2(1.4)   | 09/90 (Hd,0.7); 08/90 (CA,3.5) |
| 4C 56.27     | 18.5 | 0.664 | 20    | 5.18         | +     | 45.3(1.3)              | 05/93 (CA,2.2) |
| PKS 2005-489 | 13.5 | 0.071 | 42    | 5.08         | +     | 10.1(1.0)              | 06/90 (ESO,1.5) |
| S5 2007+777  | 16.5 | 0.342 | 24;27 | 3.02; 13.08  | + ; + | 26.9(0.9); 6.2(2.5)    | 10/91 (Hd,0.7); 08/90 (CA,3.5) |
| PKS 2131-021 | 20.0 | 0.557 | 9     | 2.40         | -     | 6.2(2.0)               | 09/93 (ESO,1.5) |
| BL Lac       | 14.0 | 0.069 | 59    | 3.37         | +     | 21.1(1.0)              | 08/90 (CA,3.5) |
| PKS 2240-260 | 17.5 | 0.774 | 41    | 6.39         | +     | 5.7(1.0)               | 09/92 (ESO,1.5) |
| PKS 2254+074 | 17.0 | 0.190 | 17;10 | 12.14; 2.28  | + ; + | 24.1(1.2); 5.3(2.0)    | 09/93 (ESO),1.5); 10/91 (Hd, 0.7) |

Column 1 gives the name of the sources, column 2 the average brightness ($m_R$) during the observations and column 3 the redshift. In column 4 the number of observations are listed, in column 5 the maximum time separation between two observations and in column 6 the result of the $\chi^2$-test. A + sign denotes the variable, a - sign the non-variable sources (see text for details). In column 7 the variability amplitude as defined in the text is given. Here the 1 $\sigma$ errors are included in brackets. Column 8 finally gives the observing dates, the observatory and the telescope used. Hd,0.7 = Heidelberg, 0.7m; ESO,1.5 = ESO Danish 1.5m; CA,3.5, 2.2, 1.2 = Calar Alto 3.5m, 2.2m and 1.2m, respectively.

classified as variable on time-scales of days or less. No BL Lac object changed its variability behaviour.

7 BL Lac objects were observed on two campaigns separated by at least 1 month (see table 2, column 5). 6 were classified as variable, 1 as non-variable. None of them changed its variability behaviour between the observations, irrespective of the confidence level of 99.5% or 90% used in the $\chi^2$-test.

The non-variable objects were in general those, which are among the faintest objects during the observations. Some of them were not observed very frequently, sometimes once per night and they have in general the largest error. Since variability on short time-scales seems to be a very common phenomenon, we tested whether we may have missed variability in the quiescent objects due to the high error or whether variability on short time-scales is related to the average brightness of the objects. We calculated the variability amplitudes $Amp = \sqrt{(A_{max} - A_{min})^2 - 2\sigma^2}$ where $A_{max}$ and $A_{min}$ are the maximum and minimum values of each lightcurve and $\sigma$ the measurement errors, respectively. The resulting distribution of the amplitudes is shown in figure 2a, the amplitudes (and the $1\sigma$ errors) for each object are given in table 2 in column 7. It can be seen that the distribution of the amplitudes is rather broad. The mean amplitude of the variable objects is 28.2±19.7%, the median amplitude is 22.5%. 4 non-variable objects have amplitudes less than 7.5% with an error below 2%, the remaining 2 displayed amplitudes between 10% and 15%, but have the highest measurement error among our observations (3.2% and 5.6%, respectively). Therefore we may have missed variability in these two BL Lac objects, but this would not change the variability behaviour of our sample in general.

In figure 2b, we show the correlation of the average brightness of the BL Lac objects during the observations with the variability amplitudes. There is a slight trend for stronger variability amplitudes towards increasing brightness. However, to prove this trend, variability measurements of the BL Lac objects from this sample during various brightness levels are necessary.

Summarizing our observations we conclude that at least 82% of the BL Lac objects from the 1 Jy sample displayed variability during the observations.

## 5. Temporal characteristics of the variability

We studied the temporal characteristics by means of statistical methods. Structure function analysis is a powerful tool to measure time-scales, especially when the data spacing is rather inhomogeneous. The results are compared to an autocorrelation analysis, which we are using to derive typical amplitudes per time interval. Additionally, this method allows to check for preferred time-scales.

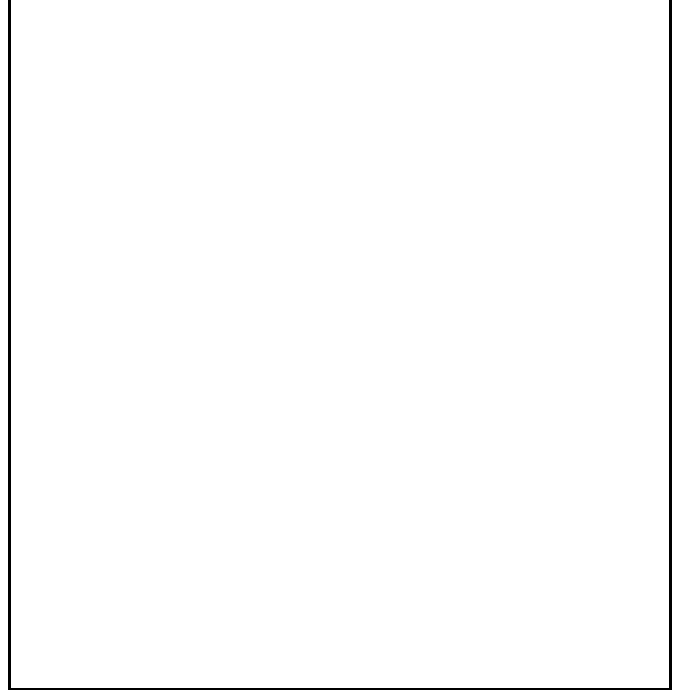

**Fig. 2.** a) Histogram of the variability amplitudes as defined in the text. The open regions correspond to the variable sources, the dotted regions to the non-variable sources and the hatched regions to the sources, which were observed twice. b) Variability amplitudes versus average brightness. The circles correspond to the variable, the dots to the non-variable sources.

### 5.1. Structure function analysis

According to Simonetti et al. (1985), we computed the first order structure function (SF) from the lightcurves:

$$SF(\Delta t) = \frac{1}{N} \sum_{i=1}^{N} (f(t_i + \Delta t) - f(t_i))^2 \quad (1)$$

The slope $\alpha$ of the SF in $\log(SF)$-$\log(\Delta t)$ space characterizes the variability and hence the underlying process. If $\alpha = 1$ shot noise dominates, towards flatter slopes ($\alpha = 0$) flicker noise becomes more important. Characteristic time-scales show up as maxima in the SF. When a lightcurve contains cyclic signals with a period P, the SF will fall to a minimum at $\Delta t = P$ (Smith et al. 1993).

First we determined the minimum time-scale, which we could resolve from the lightcurves and the maximum measurable time-scale. Following Wagner et al. (submitted) we set the shortest resolvable time-scale within a lightcurve from the time interval, where the amplitudes of the variations are larger than the measurement error. This time interval can be found by computing the absolute values of the intensity differences $\Delta I$ of two measurements as a function of time separation $\Delta t$. This is shown for S4 0954+658 in figure 3. At $log\Delta t = -1.6$, i.e. 35 minutes, error. The longest time-scale which can be measured is given by the length of the data train for each BL Lac object. In figure 4 we present the distribution the minimum and maximum time-scales for all variable BL Lac objects. 75% of all variable BL Lac objects varied significantly on time-scales shorter than 12 hours and 65% of these BL Lac objects were observed for at least 5 nights. For a homogeneous analysis we studied the properties of $SF(\Delta t)$ within $0.5\ days < \Delta t < 5\ days$.

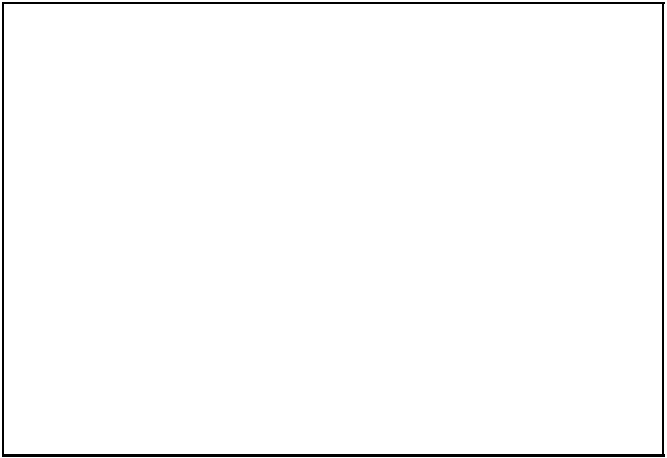

**Fig. 3.** Absolute values of all intensity differences $\Delta I$ as a function of time separation $\Delta t$ of S4 0954+658.

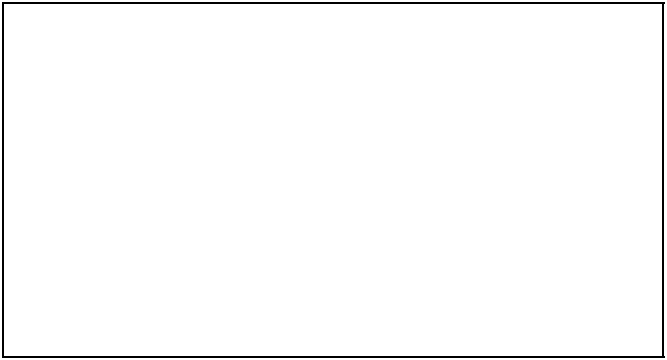

**Fig. 4.** Distribution of the shortest resolvable time-scales (dashed) and the largest measurable time-scales (solid). The vertical lines mark the time interval between 0.5 and 5 days.

Since the gaps introduced by the daylight periods are not small compared to the total length of the individual data train we did not use equidistant steps in time to compute the SF. Instead we binned the SF in variable intervals such that each interval contains the same amount depend on the total number of data points n with:

$$\Delta N = \begin{cases} 1/10 \times n(n-1)/2 & n < 30 \\ 1/20 \times n(n-1)/2 & n > 30 \end{cases} \quad (2)$$

The SF was computed for all lightcurves of the variable BL Lac objects from $\Delta t_{min} \rightarrow \Delta t_{max}$ and vice versa. The deviation of the two SF from each other gives an estimate for the statistical error. For further analysis the SF in both directions were averaged. When the last bin in both time directions contained less differences $\Delta I$ than the other bins it was excluded. In figure 5 we display the SF of the variable BL Lac objects.

### 5.1.1. The slopes of the SF

The slopes $\alpha$ of the SF were measured by fitting a power law using the least-squares method to the SF in the $\log(SF)$-$\log(\Delta t)$ space. We measured slopes only in those SFs, where the SF was determined by at least 3 bins in the range between 0.5 and 5 days. Hence we could measure slopes for 32 SFs (excluding PKS 0820+225 and MRK 501). The distribution of slopes is shown in figure 6a and given for each variable object in table 3. They range between 0 and 2.5 with a mean value of 0.8 and a dispersion of 0.6. Hufnagel & Bregman (1992) found from optical variations on time-scales from months to years of 3 BL Lac objects (PKS 0735+178, OJ 287 and BL Lac) slopes around 0.35. All these BL Lac objects are included in our sample as well. We measured a slope of 2.3, 1.6 and 0.54 for these sources indicating a steepening of the SF on a time-scale of about one week.

In the radio domain slopes of the SF were determined from variations on time-scales of month to weeks from radio monitoring programs in Metsähovi and Michigan. At 22 and 37GHz Lainela & Valtaoja (1993) derived a mean value of 1.2 for a sample of 14 BL Lacs, whereas Hughes et al. (1992) determined an average slope of 0.95 for 20 BL Lac objects at 4.8, 8 and 14.5GHz.

### 5.1.2. The characteristic time-scales determined from the SF

The characteristic time-scales of the variability were measured directly from the SF by identifying the maxima in the $log(\Delta t) - log(SF)$ plane. In 21 of the 28 variable BL Lac objects we were able to measure a time-scale. The remaining 7 BL Lac objects were classified as variable according to the $\chi^2$-test, but displayed no pronounced time-scale. For S4 0954+658, which showed maxima at 1.4, 2.6 and 6.7 days, we used the average of 3.6 days for further analysis. The resulting distribution of the intrinsic variability time-scales found is shown in figure 6b. The distribution is homogeneous between 0.5 and 3 days.

Quirrenbach et al. (1992) computed SF from intra-day variability measurements in the radio domain of flat-spectrum radio sources. They distinguished between sources with SF of type I (those with a monotonically increasing SF) and SF with type II (those with a SF showing a pronounced maximum). Given the variance of the SF in different campaigns of those objects which have been measured repeatedly we observe no clear classification in type I and II.

**Fig. 5.** SF of all variable BL Lac objects from $\Delta t_{min} \rightarrow \Delta t_{max}$ and vice versa.

### 5.2. Autocorrelation function analysis

The autocorrelation function (ACF) of each variable BL Lac object was computed using the discrete correlation function method described by Edelson & Krolik (1988). This method is especially favorable in cases where the data sampling is inhomogeneous, because no interpolation is necessary and the problem of spurious correlations at zero lag due to correlated errors is avoided. Again, as in the case of the structure function analysis, a critical point is the choice of the best binsize $\Delta t$. This depends on the evenness of the measurements between and during the nights. Following Borgeest & Schramm (1994) we defined a measure of the evenness of the distribution of observing epochs

$$\hat{n} = \frac{(t_n - t_1)^2}{\sum_{i=1}^{n-1}(t_{i+1} - t_i)^2} + 1 \qquad (3)$$

where $n$ is the number of measurements and $t_i$ are the dates of the observations. For equally spaced measurements n equals $\hat{n}$. Thus we can define a parameter $\tilde{n}$, which give the relationship between the eveness of the measurements between and during the nights

$$\tilde{n} = \frac{1}{n} \frac{\hat{n}_{(t_{i+1}-t_i > 0.5\,\mathrm{days})}}{\hat{n}_{(t_{i+1}-t_i < 0.5\,\mathrm{days})}} \qquad (4)$$

We computed $\tilde{n}$ for all variable BL Lac objects. For $\tilde{n} > 8$ we chose $\Delta t = 0.5$ days, otherwise $\Delta t = 0.2$ days.

| Object | Slope $\alpha$ | $\Delta t$ [days] | $\dot{I}$ [%/day] |
|---|---|---|---|
| PKS 0048-097 | 0.75(.09) | 1.9(.20) | 0.35(.34) |
| AO 0235+164 | 0.57(.23) | | 6.33(3.14) |
| PKS 0426-380 | 0.03(.07) | 2.5(.30) | 20.55(1.81) |
| S5 0454+844 | 2.37(.34) | 3.1(.37) | 1.93(1.53) |
| S5 0454+844 | 0.75(.16) | | |
| PKS 0537-441 | 0.49(.11) | 2.1(.33) | 4.42(1.78) |
| S5 0716+714 | 0.01(.06) | 1.5(.16) | 8.83(4.23) |
| PKS 0735+178 | 2.28(.91) | 2.4(.28) | |
| OJ 425 | 0.77(.58) | 2.4(.24) | 0.82(.96) |
| PKS 0820+225 | | | 2.25(.84) |
| OJ 287 | 1.60(.19) | 2.8(.30) | 4.51(3.45) |
| S4 0954+658 | 0.76(.14) | 2.6(.55) | 2.69(2.2) |
| PKS 1144-379 | 0.42(.15) | 0.5(.08) | 26.92(3.44) |
| B2 1147+245 | 1.77(.49) | 0.8(.08) | 0.17(1.17) |
| B2 1308+326 | 0.07(.22) | | |
| OQ 530 | 1.12(.18) | 2.5(.24) | 3.00(3.00) |
| AP Lib | 1.17(.11) | | 0.69(.99) |
| 4C 14.6 | 0.97(.09) | 1.3(.17) | 2.26(1.23) |
| MRK 501 | | | 0.33(.71) |
| OT 081 | 1.93(.07) | | 15.99(5.91) |
| S4 1749+701 | 1.11(.19) | 1.0(0.3) | 11.32(4.48) |
| S4 1749+701 | 0.20(.76) | | |
| S5 1803+784 | 0.83(.21) | 1.1(.16) | 12.64(4.19) |
| S5 1803+784 | 0.58(.13) | | |
| 3C 371 | 0.48(.25) | 2.4(.20) | 0.58(3.56) |
| 3C 371 | 0.16(.32) | | |
| 4C 56.27 | 1.40(.08) | | 6.55(1.81) |
| PKS 2005-489 | 0.91(.12) | 0.6(.05) | 0.23(.57) |
| S5 2007+777 | 0.73(.22) | 0.6(.07) | 2.07(1.32) |
| S5 2007+777 | 0.51(.16) | 0.6(.07) | |
| BL Lac | 0.54(.27) | 1.7(.15) | 2.69(3.02) |
| PKS 2240-260 | 0.62(.08) | 2.89.41) | 1.28(0.63) |
| PKS 2254+074 | 0.03(.31) | 1.3(.13) | 0.61(1.43) |
| PKS 2254+074 | 0.17(.05) | | |

**Table 3.** Results of the statistical analyses of the variable BL Lac objects. Column 1 gives the name of the sources, column 2 the slopes $\alpha$ of the SF, column 3 the time-scales found from the SF and column 4 the activity parameter $\dot{I}$ measured from the ACF.

### 5.2.1. The activity parameter $\dot{I}$

In order to measure a typical amplitude per time interval of a BL Lac object we computed the activity parameter $\dot{I} = \sqrt{ACF(0)}/\Delta t$ from the ACF. Therefore we did not normalize the ACF to unity at $\Delta t = 0$. We fitted a second-order polynomial to each ACF. ACF(0) was derived from the maximum of the fit polynomial, for $\Delta t$ we chose the decorrelation length of the fit polynomial. With two exceptions (PKS 0735+178 and B2 1308+326) we could mea-sure $\dot{I}$ for all variable BL Lac objects. The values were corrected for redshift. Their distribution (given in %/day) is shown in figure 7, the value for each BL Lac object is given in table 3. 16 out of 26 BL Lac objects displayed a moderate activity of 3%/day or less. The remaining 10 BL Lac objects showed stronger activity up to 27%/day.

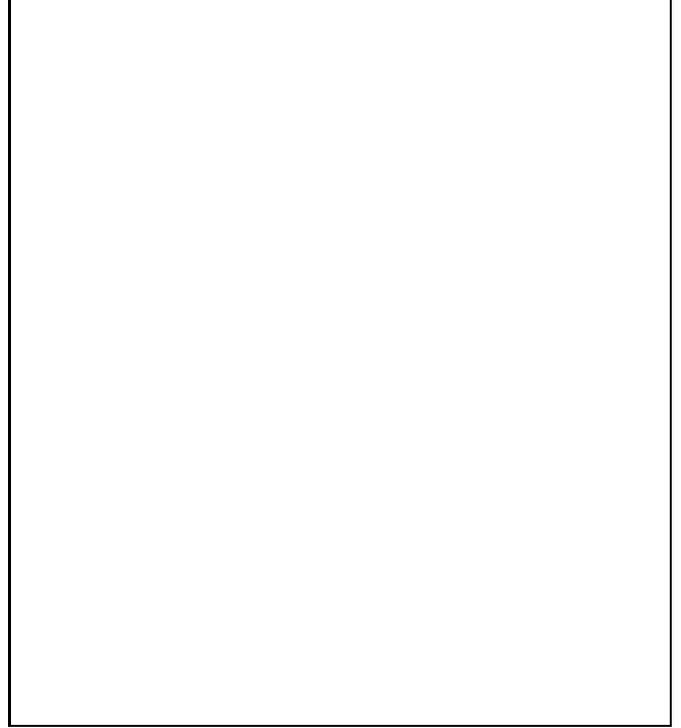

**Fig. 6.** a) Distribution of slopes $\alpha$ from the SF. b) Distribution of the intrinsic time-scales found from the SF

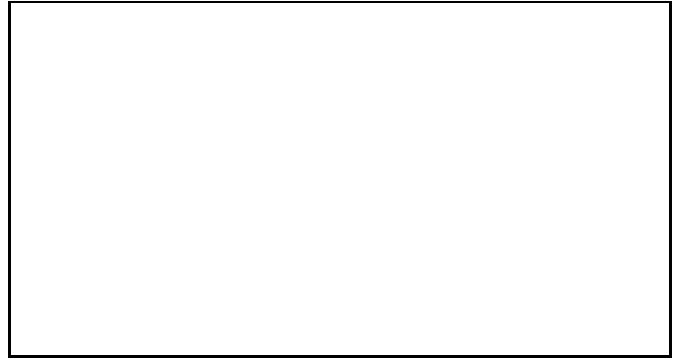

**Fig. 7.** Distribution of the activity parameter $\dot{I}$ measured from the ACF.

Quasiperiodic oscillations on short time-scales have been observed only in a few BL Lac objects so far. For example, Wagner (1991) found in S5 0716+714 quasiperiodic variations on time-scales of 1 and 7 days, respectively. These variations were observed simultaneous in the radio and the optical domain. Urry et al. (1993) found quasiperiodic oscillations in PKS 2155-304 both in the UV and X-ray regime on time-scales of one day. Such periodicities in the lightcurves can be found by searching for harmonic components in the ACF. We checked this for all variable BL Lac objects and found in S5 0716+714 a harmonic component with a period of 4 days. The ACF is shown in figure 8. For comparison an ACF of a variable but non-periodic BL Lac object is shown, too. The maxima (and minima) separated by 4 days can clearly be seen. Minor maxima show up in the ACF at time-scales of one day as well, but this time-scale is strongly influenced by the day/night change and cannot be determined unambiguously. In the remaining ACF no periodicities were found.

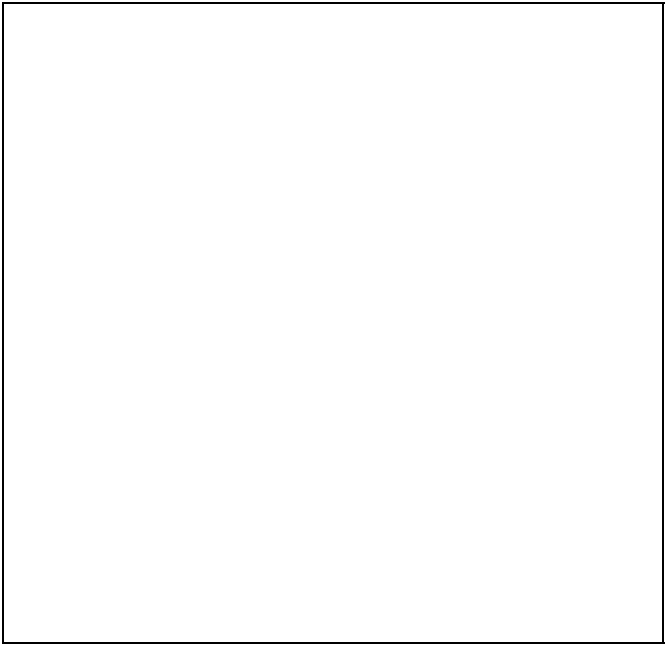

**Fig. 8.** ACF of S5 0716+714 and S4 0954+658. The error bars correspond to 1 $\sigma$ errors. The individual points were combined to trace the eye.

## 6. Discussion

Although variability is one of the defining criteria of BL Lac objects, no complete sample has previously been investigated for the characteristics of optical variability on time-scales of days. 28 of the 34 BL Lac objects from the 1 Jy sample displayed intraday variability. This corresponds were observed twice or more often. None of them changed its variability behaviour. Again the fraction of variable objects was higher than 80%. Assuming that this detection rate could be transformed in a duty cycle (the fraction of time, where a BL Lac object is variable), the duty cycle in this sample would be at least 0.8.

A comparison with another variability study show a similar trend. Wagner et al. (1990) searched for intraday variability in a small subset of the 1 Jy sample. They detected intraday variability in 3 out of 4 BL Lac objects corresponding to a fraction of 75%.

In 21 of the 28 variable BL Lac objects we could measure a time-scale which lies in the range of 0.5-3 days. This demonstrates again that intraday variability is a characteristic property of BL Lac objects from the 1 Jansky sample. Whether this is a characteristic property of the whole BL Lac class or whether there are intrinsic differences between the RBL and the XBL is still unclear (Heidt et al. in preparation).

We checked, if there are any dependences of the time-scales on intrinsic properties, redshift or absolute magnitude and found none. Clearly, these observations have to be confirmed by repeated intraday variability studies of the 1 Jy sample.

The activity parameter $\dot{I}$ shows a bimodal distribution. 10 BL Lac objects displayed an $\dot{I}$ between 3 and 27%/day, the remaining sources less than 3%/day. We correlated $\dot{I}$ with the redshift and the absolute magnitude, which we calculated roughly from our observations. Figure 9 displays $\dot{I}$ as a function of $M_R$ and z.

The high-redshift BL Lac objects have a higher $\dot{I}$ on average and a higher dispersion among the $\dot{I}$ than the low-redshift BL Lac objects. The mean $\dot{I}$ for the BL Lac objects in the range $0 < z < 0.4$ is 2.9±4.3 %/day, for the BL Lac objects with a redshift $z > 0.6$ $\dot{I}$ is 7.5±7.8 %/day. This result is not in disagreement with suggestions by Burbidge & Hewitt (1991) and Valtaoja et al. (1991). Burbidge & Hewitt (1991) proposed to use the term BL Lac for BL Lac candidates only, when their spectra are typical for non-thermal sources lying in bright elliptical galaxies. According to their definition all BL Lac objects detected so far have a redshift $z \leq 0.6$. They proposed to use the term strong variable quasars for those BL Lac candidates, whose spectra have either broad emission lines or absorption spectra of QSO type. A similar suggestion was proposed by Valtaoja et al. (1991) by means of variability studies in the radio domain. They distinguish between close BL Lac objects ($z < 0.3$) with low Doppler factors ($\sim 1$) and distant BL Lac objects ($z > 0.3$) with high Doppler factors ($\sim 1...8$). They conclude that the parent population of close BL Lac objects are FR I radio galaxies and the parent population of distant BL Lac objects are FR II radio galaxies. Both suggestions would propose that in distant BL Lac objects $\dot{I}$ should be higher on average, what we observed indeed.

**Fig. 9.** Activity parameter $\dot{I}$ versus absolute magnitude $M_R$ (upper panel) and activity parameter $\dot{I}$ versus redshift (lower panel). The dots correspond to BL Lac objects whose redshift was measured either from emission lines or absorption lines of their host galaxy or given as lower limits in Stickel et al. (1993), the circles to BL Lac objects, where no redshift information is available. Note that the time-scales involved in the derivation of $\dot{I}$ have been corrected for cosmological effects.

These suggestions are challenged by VLBI polarization measurements, which show that the magnetic fields in jets of BL Lac objects are preferably oriented perpendicular to the jet axis, whereas they are in quasars and radio galaxies oriented parallel along the jet axis (Wardle et al. 1986; Gabuzda et al. 1989, 1991; Cawthorne et al. 1993a, b).

We found a trend for higher $\dot{I}$ towards higher $M_R$. This may be due to relativistic beaming. Curiously, the BL Lac object with the highest $M_R$ during our observations (PKS 0537-441) displayed only a "moderate" $\dot{I}$. This BL Lac has one of the highest historical amplitudes of all BL Lac objects in the 1 Jy sample ($\Delta m_B = 5.4$mag, Usher, 1975). In earlier measurements it varied by more than two magnitudes within 2 month (Eggen, 1973). Although it has pears surrounded by nebulous filaments on deep images (Stickel et al. 1988). One suggestion is that this structure belongs to a galaxy along the line-of-sight with a redshift around z = 0.2. Hence the chance is high that the radiation from the BL Lac is enhanced by gravitational microlensing (Stickel 1991). Within this framework our observed properties could be a superposition of intrinsic variability (which causes the short time-scale variations) and gravitational microlensing (which causes the high $M_R$). Contradictionary, Falomo et al. (1992) failed to detect the presence of a foreground galaxy and Narayan & Schneider (1990) argued based on theoretical arguments that the assumed redshift of the intervening galaxy may be in error.

Our observations can be described by the standard model of shocks moving down a relativistically beamed jet as proposed by Marscher & Gear (1985). The time-scales observed are close to the lower limit of this model (1 day). Shorter time-scales may still be present in the lightcurves, but in most cases the sampling is not dense enough to resolve them. Nevertheless there are a few exceptions, where modifications of this model or alternative explanations should be taken into account. The BL Lac S4 0954+658 showed in February 1990 several symmetrical outbursts lasting 2-3 days with amplitudes up to several 100% (Wagner et al. 1993). It was observed by us again in 1991 and 1992. As in 1990, we observed strong symmetrical outbursts on similar time-scales. Furthermore we found nearly identical flares separated by one year. This is demonstrated in figure 10, where the lightcurve of S4 0954+658 from January 1991 is shown. Overlayed is a flare, which we observed in January 1992. All these observations suggest that in this BL Lac geometrical effects may play an important role, e.g. knots floating within a relativistic, rotating jet as proposed by Camenzind & Krockenberger (1992) and already successful applied to outbursts of 3C 345 in 1990/1991 (Schramm et al. 1993) and PKS 0420-014 in 1990-1992 (Wagner et al. 1995).

In S5 0716+714 we found a quasi-periodic signal with a time-scale of 4 days, during observations in February 1990 quasiperiodicities on time-scales of 1 and 7 days were found, respectively (Quirrenbach et al. 1991). Since then this BL Lac was observed regularly by us. In most cases quasi-periodic oscillations were found (Heidt 1994). Within the shock-in-jet model, it is not easy to assume a scenario, where shocks within the jet are injected with a constant rate and thus show up as periodicities in the lightcurve. In an alternative model periodic signals can be modelled by "hot spots" rotating within the accretion disk of an AGN (Abramowicz et al. 1991, Mangalam & Wiita 1993). However, to explain the observations in S5 0716+714 with this model just one "hot spot" could rotate at a specific time otherwise the periodicities would be washed out. Variability related to accretion disks would occur in radio-quiet objects. Up to now intraday variabil-ity has not been found in such QSOs (Gopal-Krishna et al. 1993a, b).

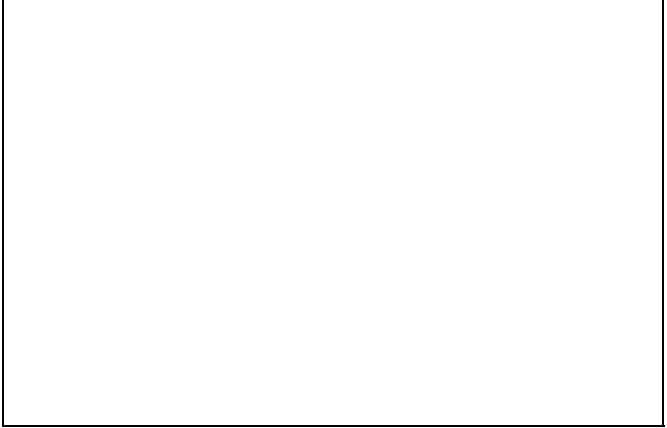

**Fig. 10.** Lightcurve of S4 0954+658 in January 1991 (circles). Overlayed observations from January 1992 (dots). Both lightcurves were scaled in such a way that the relative scaling is equal. Note the perfect match between both flares around JD 2448266/2448641 (taken from Heidt, 1993).

## 7. Conclusions

We have studied the variability behaviour of a complete sample of radio-selected BL Lac objects taken from the 1 Jy catalogue on short time-scales. Our results can be summarized as follows:

1) In 28 out of 34 BL Lac objects (82%) we detected variability on time-scales of days. Similar detection rates have been found in repeated measurements and other studies. This implies that the duty cycle in these objects is very high, i.e. at least 0.8.

2) 21 BL Lac objects show a characteristic time-scale in the range of 0.5-3 days. We found no correlation of the time-scales with either the redshift or the absolute magnitude.

3) We investigated an activity parameter $\dot{I}$ from the variable BL Lac objects. The distribution is bimodal. 16 (62%) of the BL Lacs displayed an $\dot{I}$ of less than 3%/day. The remaining 10 BL Lacs showed an $\dot{I}$ of up to 27%/day. On average $\dot{I}$ is higher in high redshift BL Lacs. Furthermore we found a tendency of higher $\dot{I}$ towards higher absolute magnitudes. This may be due to relativistic beaming. In the BL Lac with the highest absolute magnitude we found only a moderate $\dot{I}$.

4) The observations can in general be explained with the standard shock-in-jet model. However, in one BL Lac (S5 0716+714) we found quasi-periodicities, which can hardly be explained by the standard model. Another BL Lac (S4 0954+648) displayed a variability behaviour which demonstrated that at least in this BL Lac geometrical effects could play an important role. These two BL Lac obstrates that the various models can only adequate be tested with long and densely sampled lightcurves. In any case, all models, which are able to explain variability on time-scales of days must be able to consider the high duty cycle of these BL Lac objects.

*Acknowledgements.* The authors would like to thank U. Erkens for assistance during the observations and Drs. A. Witzel and M. Camenzind for valuable and stimulating discussions. This work was supported by the DFG (Sonderforschungsbereich 328).